\documentclass[sigconf]{acmart}
\usepackage{colortbl}
\newcolumntype{r}{>{\columncolor[RGB]{240,240,255}} p{1.2cm}}

\AtBeginDocument{%
  \providecommand\BibTeX{{%
    \normalfont B\kern-0.5em{\scshape i\kern-0.25em b}\kern-0.8em\TeX}}}

\copyrightyear{2021}
\acmYear{2021}
\setcopyright{rightsretained}
\acmConference[CHI '21]{CHI Conference on Human Factors in Computing Systems}{May 8--13, 2021}{Yokohama, Japan}
\acmBooktitle{CHI Conference on Human Factors in Computing Systems (CHI '21), May 8--13, 2021, Yokohama, Japan}\acmDOI{10.1145/3411764.3445626}
\acmISBN{978-1-4503-8096-6/21/05}

\begin{document}

\title{TapNet: The Design, Training, Implementation, and Applications of a Multi-Task Learning CNN for Off-Screen Mobile Input}
\author{Michael Xuelin Huang, Yang Li, Nazneen Nazneen, Alexander Chao, Shumin Zhai}
\affiliation{%
  \institution{Google, Inc.}
  \city{Mountain View}
  \state{CA}
  \country{USA}
}
\email{{mxhuang, liyang, fnazneen, alexanderchao, zhai}@google.com}






\renewcommand{\shortauthors}{Huang, et al.}


\begin{abstract}
 
To make off-screen interaction without specialized hardware practical, we investigate using deep learning methods to  process the common built-in IMU sensor (accelerometers and gyroscopes)  on mobile phones into a useful set of one-handed interaction events. We present the design, training, implementation and applications of TapNet, a multi-task network that detects tapping on the smartphone. With phone form factor as auxiliary information, TapNet can jointly learn from data across devices and simultaneously recognize multiple tap properties, including tap direction and tap location. We developed two datasets consisting of over 135K training samples, 38K testing samples, and 32 participants in total. Experimental evaluation demonstrated the effectiveness of the TapNet design and its significant improvement over the state of the art. Along with the datasets, codebase\footnote{\href{https://sites.google.com/site/michaelxlhuang/datasets/tapnet-dataset}{TapNet Dataset}}, and extensive experiments, TapNet establishes a new technical foundation for off-screen mobile input.


\end{abstract}




\maketitle

\section{Introduction}

\begin{figure}
\centering
  \includegraphics[width=1.\columnwidth]{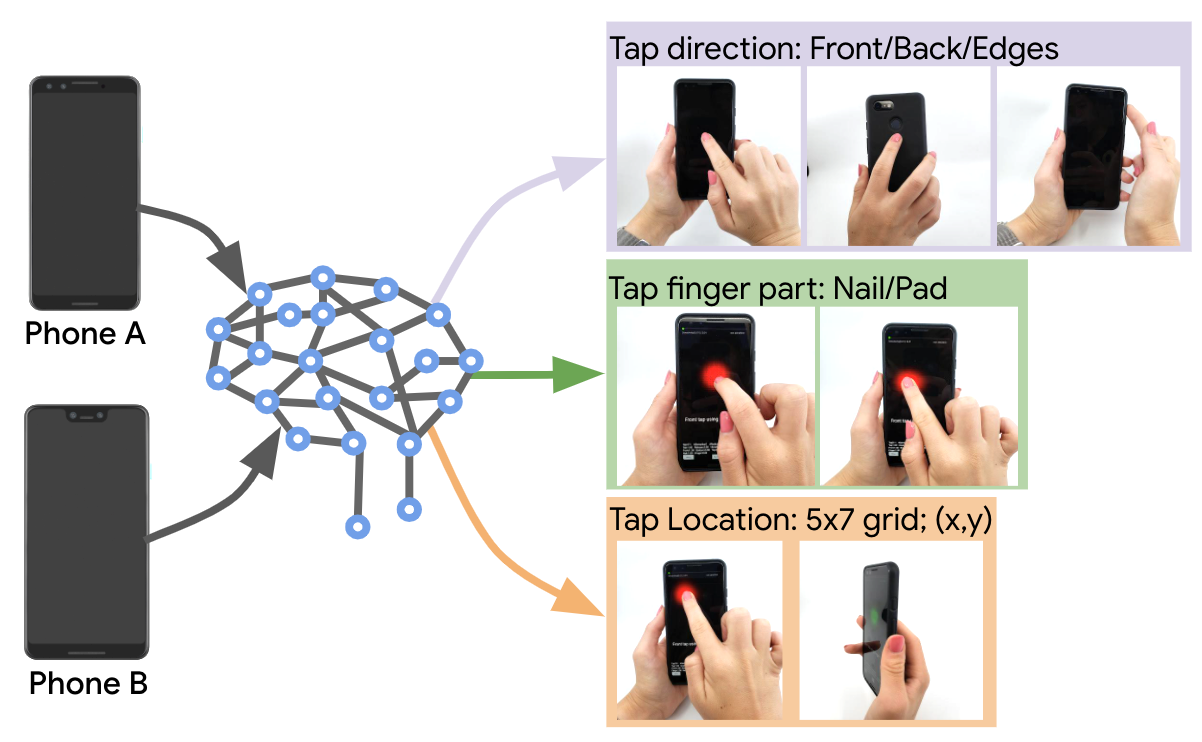}
  \caption{TapNet is a one-model-for-all solution that jointly learns from cross-device data and performs multiple tap recognition tasks at a time.}~\label{fig:teaser}
  \Description{Input of the neural network is the data of multiple phones, and its output is the tap recognition results of direction classification, finger part classification, tap location regression, and so on.}
\end{figure}

Touchscreen as the sole method of mobile device input is increasingly challenged by its inherent limitations. The difficulty of one-handed use and the visual  occlusion by the operating finger are two of them. Beyond research explorations, we began to see off-screen occlusion-free alternatives in mainstream products, such as double pressing power button to activate the camera, squeezing the Active Edge~\cite{quinn2019active} of  Google Pixel phones and long-press power button~\cite{apple} of the iPhones for voice command invocation.


In addition to a \textit{tap event} (presence of a tap) alone, potentially useful \textit{tap properties} include \textit{tap direction}~\cite{mcgrath2014detecting, xdadevelopers2020} (i.e. whether the tap is on the front, back, or side of the phone), \textit{tap location}~\cite{liang2018deep,mehrnezhad2018stealing} (i.e. which region the tap falls on the devices), and tap \textit{finger part}~\cite{harrison2011tapsense} (i.e. tapping with finger pad vs fingernail). Prior studies has focused on the recognition of a single tap property, based on a limited amount of data as a proof-of-concept. 
This paper aims to address the needs in practice, by predicting comprehensive tap properties for diverse application purposes and advancing the state of the art of off-screen interaction towards a practical level of performance. Note that the state of the art for this topic can only be advanced if there are well-established benchmarks with accessible datasets and codebases. Our work is set to help create a benchmark with extensive dataset development, neural network model design, model training, and experimentation.

After experimenting with multiple methods and architectures for tap detection, 
we found that a multi-input, multi-output (MIMO) convolutional neural network (CNN)  gave the best results. 
The resulting method, TapNet, enables joint learning on cross-device data and joint prediction of multiple tap properties.
Compared with the solution of one model for each task, a joint model (see Figure~\ref{fig:teaser}) can  exploit the interrelation among multiple tap properties during training and also saves  run-time computation and memory. 
TapNet contains shared convolutional layers for task-agnostic knowledge extraction, meanwhile each of its output branch retains the task-specific information. 
TapNet uses the inertial measurement unit (IMU) signals as primary input and the phone form factor as ``auxiliary information''~\cite{stephenson2004speech} for \textit{cross-device training}, i.e. joint training on cross-device data.
Together, TapNet offers a one-model-for-all solution~\cite{kaiser2017one} across devices. 

Our technical investigation and evaluation shed lights on two important aspects for machine learning (ML) in HCI development: methods and data.
First, thanks to the MIMO architecture, TapNet increased tap property recognition accuracy over the prior art methods, particularly for the more difficult tasks such as tap location recognition. 
Specifically, TapNet yields a mean distance error around 10\% of the screen diagonal, similar to the between-icon distance on mobile phone home page.
Although this resolution is much less precise compared to the touchscreen, such resolution can already enlarge the interaction design space, by for example enabling users to perform selection by tapping on the back of the phone, even when they wear gloves.

Second, we discuss an efficient data strategy for developing a ML based HCI application.
Different from most computer vision tasks that heavily rely on multi-person data to achieve user generalizability~\cite{huang2018quick,zhang2017mpiigaze}, a well-performing ML model for HCI (interactive) tasks may not necessarily demand multi-person training data. 
The key is to ensure the training data diversity.
To test this hypothesis, we collected a one-person dataset for training and a separate multi-person dataset for  testing (and optionally model adaptation). 
The results show that the one-person model could achieve a comparable level of user generalizability as a model tuned on multi-person data.


Our contribution is four-fold.
We 1) developed a multi-task network that can be jointly trained across devices. The network could simultaneously recognize  a set of tap properties, including tap direction and location;
2) developed two datasets and conducted extensive evaluations that advanced the state of the art;
3) offered new perspectives on  alleviating the data hurdle in ML based HCI research;
4) established a benchmark with opensource code and datasets for off-screen tapping recognition, which will be reproducible and accessible by others.


\section{Related work}

This study is related to off-screen interaction, in particular, tap-based interaction, as well as multi-task neural networks~\cite{caruana1997multitask}.

\subsection{Off-Screen Interaction}

Back-of-device (BoD)~\cite{corsten2017backxpress,de2013back,le2016investigating,le2018infinitouch} and edge-of-device interactions~\cite{holman2013unifone,le2018infinitouch,yeo2016sidetap,zhang2016watchout} have attracted much attention, however, most of them relied on specialized sensors that are not readily available on phones.
For instance, BackXPress studied finger pressure for BoD interaction using a sandwiched smartphone~\cite{corsten2017backxpress}.
InfiniTouch recognized finger location from capacitive image outside the touchscreen~\cite{le2018infinitouch}. 
Some exploited small widgets to enable off-screen gesture sensing.
Wearing a magnet can allow magnetometer to track the 3D finger movement~\cite{chen2016finexus,reyes2018synchrowatch}.
Adding a mirror can make camera to detect finger location on the back~\cite{wong2016back} or around the device~\cite{yu2019handsee}.
Acoustic sensing exploits sound propagation properties  to recognize BoD gesture~\cite{sun2018vskin}, grip force~\cite{tung2016expansion}, and contact finger part~\cite{harrison2011tapsense}, and electric field sensing  detects around-device gestures in a non-intrusive manner~\cite{zhang2017electrick,zhou2016aurasense}. 
Unlike using the IMU, these detection methods require additional hardware installed on the already very compact mobile devices, increasing manufacturing and material cost and potentially reducing available space to the largest possible battery. 

\subsection{Tap Detection from IMU Signals}

Despite the line of research on detecting tapping from motion signals captured by the built-in IMU sensors on smartphones ~\cite{granell2016less,seipp2014backpat,zhang2016beyond,zhang2013backtap}, there is room for improvement.
SecTap allowed users to move a cursor by tilting and select by back tapping~\cite{ling2018sectap}.
Bezel-Tap detected tap on the edge of the devices~\cite{serrano2013bezel}.
Granell and Leiva conducted feature engineering for tap detection~\cite{granell2016less}.
In contrast to the detection of tap event alone, BackPat recognized tapping in three locations based on gyroscope and microphone signals~\cite{seipp2014backpat}.
BackTap~\cite{zhang2013backtap} and BeyondTouch~\cite{zhang2016beyond} classified four-corner tap based on accelerometer, gyroscope, and microphone signals. 
These studies relied on simple features (e.g. mean, kurtosis, and skewness) and statistical methods typically based on shallow neural networks ~\cite{granell2016less,ling2018sectap,seipp2014backpat,zhang2013backtap,zhang2016beyond}, thus only achieved limited precision of tap location detection.  

More recently, researchers started to explore neural networks for tap location classification for PIN code inference in the field of privacy and security~\cite{liang2018deep,mehrnezhad2018stealing}.
PINlogger classified tapping on ten buttons using a single layer neutral network~\cite{mehrnezhad2018stealing}.
Liang et al. applied a two-layer CNN model to estimate tap location from z-axis signal of accelerometer~\cite{liang2018deep}.
Although these studies yielded promising result, their network capacity was relatively small and the reported accuracy was still far from practical level. 

\subsection{Multi-Task Learning and Training with Auxiliary Information}

To improve tap recognition, we develop a multi-input and multi-output a neural network that can estimate multiple tap properties. 
This section reviews  two related concepts: multi-task learning~\cite{caruana1997multitask} and learning with auxiliary information.
Multi-task learning with neural network leads to a multi-branch architecture. 
Each branch addresses one of the recognition tasks, such as different targets in tracking~\cite{nam2016learning},  languages in translation~\cite{kaiser2017one}, and head poses in gaze estimation~\cite{zhang2018training}. 
The shared layers of all branches extract the common knowledge across tasks, whose generalizability is ensured by the regularization effect across tasks. Multi-task models employ multiple loss functions thus more supervisions during training that can potentially helps learning. 
To our knowledge, no attempt has been made to apply multi-task learning for off-screen tap recognition and the multi-task architecture can conduce to the learning of each tap recognition task from their interrelationship. 

We are also the first to exploit the form factor of the phone as auxiliary information, which has been shown to be beneficial for training.
Zhang et al. investigated the benefit of using auxiliary information in training and found that it can improve performance in testing even without using the auxiliary information as input~\cite{zhang2014can}.
Stephenson et al. pointed out that conditioning on auxiliary information can achieve higher robustness than that of appending auxiliary information directly to the main features~\cite{stephenson2004speech}.
Liao et al. showed that integrating simple but essential auxiliary information can increase prediction accuracy~\cite{liao2018deep}.

\begin{figure*}[t]
\centering
  \includegraphics[width=\textwidth]{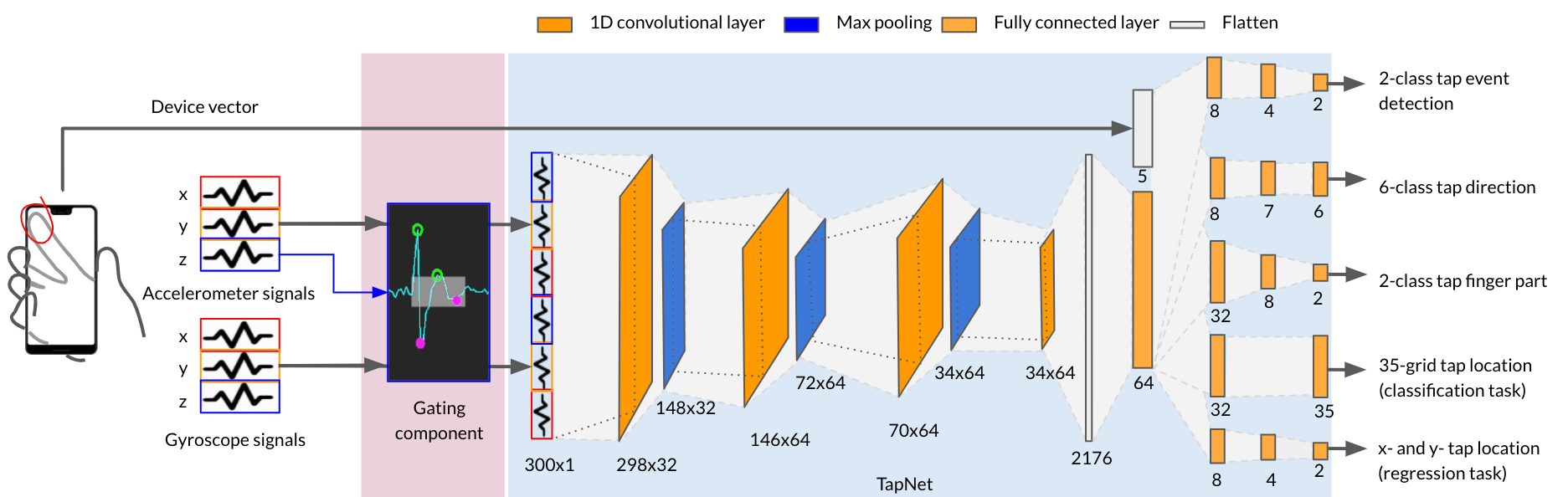}
  \caption{Accelerometer and gyroscope signals are processed in the Gating component and passed to TapNet if met set criteria. Using phone form factor (embedded in the device vector) as auxiliary information, TapNet jointly recognizes multiple tap properties, including tap event, direction, finger part, and location. }~\label{fig:architecture}
  \Description{A figure of system overview.}
\end{figure*}

\section{Recognizing Tap Properties}

Moving beyond the prior work in this space, this project aims at developing IMU-based input methods that meet the requirement of practical applications, by means of deep neural network design and training. The  key objective is to achieve the five recognition tasks (i.e. the five network outputs) with one network (TapNet) as shown in Table~\ref{tab:attribute}.
Each \textit{task} aims to recognize one \textit{tap property}, such as direction and location.
We first describe the pipeline overview, followed by the core of the method (a MIMO network). We then discuss the gating component and  signal filtering for  recognition.

\begin{table}[h]
  \centering
  \begin{tabular}{ p{1.3cm}||p{1.3cm}||p{4.5cm}}
    \toprule
    {\small \textit{Task}} & {\small  \textit{Paradigm}}&{\small  \textit{TapNet Output}}\\
    \cmidrule(r){1-3}
    {\small Event} & {\small  2-class} & {\small tap, non-tap}\\
    {\small Direction} & {\small  6-class} & {\small front, back,left, right, top, bottom}\\
    {\small Finger part} & {\small  2-class} & {\small finger pad, finger nail}\\
    {\small Location} & {\small  35-class} & {\small the region ID in a 5x7 grid}\\
    \cmidrule(r){1-3}
    {\small Location} & {\small  regression} & {\small x- and y- ratio compared with screen width and height}\\
    \bottomrule
  \end{tabular}
    \caption{The five tasks of TapNet: four classification tasks to detect different tap properties, such as tap direction and location, and one regression task to estimate tap location.}~\label{tab:attribute}
    \Description{A table of the descriptions of the five tap recognition tasks.}
\end{table}

\subsection{Pipeline Overview}

As shown in Figure~\ref{fig:architecture}, the system listens to the six-channel IMU (three-axis accelerometer and three-axis gyroscope) data and maintains a 150-ms data window.
To avoid unnecessary down-stream computation for recognizing tap properties, a heuristic-based gating mechanism using the z-axis signal of the accelerometer is performed to reject obvious non-tap motions.
If the signal passes the gating, the six-channel IMU signals within a 120-ms feature window will be concatenated into a one-dimensional feature vector.
The feature window is identified and aligned across tap samples by the tap-induced peak in the z-axis signal.
TapNet, a multi-input, multi-output convolutional neural network, takes this feature vector and a device vector as input.
The \textit{device vector} depicts the phone size, IMU pointing direction, and installation location relative to the upper left corner of the device housing.
TapNet then outputs the  predictions of five tap properties at a time as shown in Table~\ref{tab:attribute}.



\subsection{Design of TapNet Architecture} \label{Architecture}

TapNet uses a multi-input and multi-output architecture. 
We use IMU signals as primary input and device vector that describes the phone form factor as auxiliary input.
Using this auxiliary information helps to accommodate the difference of device form factor and achieves the cross-device training.

The output of TapNet contains multiple branches.
Since different tap properties have a confound impact on IMU responses, we jointly learn the mappings from IMU signals onto these interrelated tap properties using a multi-layer CNN. 
In this architecture, different tap-related tasks share four convolutional layers, which extract the common shape patterns that are indicative across tap properties. 
Following the shared layers, there are branches of fully connected layers, each of which extracts property-specific patterns for individual task. 
In practice, shared layers for tap properties also means shared computation. 
Therefore, TapNet requires less computation and memory than multiple networks each trained for a single task. 

Training  over an intersection of multiple tap tasks confines the  learning in a restrained feature embedding space and thus allows it to converge to a solution for all related tasks~\cite{kaiser2017one}.
Multi-task learning allows for good alignment between feature embeddings of different tasks. 
Such additional guidance or supervision can potentially prevent over-fitting especially given few training samples.


\subsection{One-Channel Convolutional Layers}

As illustrated in Figure~\ref{fig:architecture}, a one-channel CNN is used for tap recognition. 
We exploit convolutional layers to extract shape features from IMU signals, because the convolutional filter offers temporal locality to capture signal dynamics and shared weights to reduce trainable parameters.
In general, a convolutional filter in early layers describes the basic shape features, such as a peak, a valley, or a certain degree of slope.
A convolutional filter in later layers, with a large receptive field, is more likely to contain shape semantics, such as a large magnitude peak or an impulse with double peaks. 

Regarding the options between one-channel vs multi-channel network, we see that the one-channel network can be more efficient as it allows for filter reuse across IMU channels.
Conventional methods applied multi-channel CNN to describe signal alignment across channels, and it has been widely use to handle EEG~\cite{schirrmeister2017deep} and IMU~\cite{yang2015deep} data.
However, a large number of filters are required to depict the fine-grained signal alignment combinations of the six-channel signals, and thus increases the model training difficulty and the demand for data. 
In contrast, we propose to concatenate the six-channel data into a one-dimensional vector and apply one-channel convolutional layers in TapNet. 
By doing so, each one-channel convolutional filter can be reused to detect shape features across channels, and then rely on the fully connected layers to draw decision by associating filter activations in different parts of the signals.
Dividing the shape extraction and alignment analysis conceivably mitigates the model training difficulty and achieves an efficient use of data. 
We also evaluated this in an experiment.

\subsection{Signal Gating for Neural Network }

Although TapNet is rather lightweight compared with vision-based convolutional networks, performing recognition per frame generates unnecessary overheads.
Our observation, as well as previous findings~\cite{liang2018deep}, suggests that a peak or valley in the z-axis output is a necessary (high recall) but not sufficient (low precision) signal to a tapping event on the device.
As such, we use  the z-axis signal of accelerometer as a \textit{gating signal} to the CNN in order to minimize the amount of computation (hence power consumption) on the mobile device. 
Gating signal alone may not be enough for accurate recognition of tap properties, it is sufficient for tap-like (high recall low precision) event detection, and thus qualified as gating for the CNN recognition. 
Therefore, we only perform CNN recognition when we observe a tap-like signal from the gating signal.

Specifically, we first perform a simple, linear complexity peak detection on the gating signal at the per-frame level (Figure~\ref{fig:architecture} the pink region).
This peak detection yields a set of peaks and valleys and their corresponding timestamps.
We define \textit{a tap-like signal} as the one has an impulse that contains at least one peak, and an \textit{impulse} as a group of peaks between each pair the interval is less than the threshold, $T_v$, which is empirically set to 80 ms.
We also apply a magnitude threshold on the peak and valley to control the sensitivity of the gating component.




\subsection{Sensor Signal  and Temporal Alignment}

Sensor system on smartphones provides motion data in a number of formats, including the raw signal, its gravity excluded counterpart, and rotation vector.
As we aim to detect tap event and identify tap properties in an orientation-invariant manner, we leverage the raw accelerometer and gyroscope data and compute their first-order derivative.
Therefore, these signals represent the change of force on the housing of the phone and the resulting change of rotation velocity, and they stay at zero when the phone is stationary.
In addition to orientation invariance, we also see that the first-order derivatives accelerometer and gyroscope signals have high shape similarity.
As such, using this representation conduces to the reuse of convolutional filters and alleviates network training difficulty. 

All tap feature vectors are temporally aligned. 
Specifically, 
The first major peak in the accelerometer z-axis is aligned to a designated frame (=106th) in the feature vector, and this determines the feature window location.
Then 50 sensor samples ($\sim$120ms) from each IMU channel are concatenated to form a 300-element feature vector for each tap-like event. 
Such temporal alignment allows the network to focus on the shape differences induced by a tap.

\section{Dataset Development}

This section describes two datasets: 
the one-person dataset (135,260 samples) was used for TapNet training only (not testing) and the multi-person dataset (38,545 samples) was for performance testing and user generalizability tuning.
Our hypothesis is that \textit{a well-designed data collection protocol can ensure training data diversity even from one person and be sufficient for some tap recognition tasks}.
When we need to improve the model, it is relatively easy to enlarge the one-person training set, evaluate on the test set, and repeat this process in an iterative fashion. This is more efficient than collecting additional multi-person data.


\subsection{One-Person Training Dataset}


We used a single researcher training strategy.
One of us produced the entire training data following a  comprehensive data collection protocol, which aims to cover the diversity in real use.
Tap data were collected through multiple sessions on Phone A and Phone B\footnote{Phone A = Pixel 3 and Phone B = Pixel 3 XL}.
Each session aims to collect tap in a specific condition described by a set of characteristics of the phone, grip gestures, and the tapping itself (see Table~\ref{tab:single-person}). 
Visual indicator was presented on the screen to guide the experimenter. 
The experimenter can examine the finger alignment with the target location by turning the phone back and forth to ensure the annotation correctness.
Please refer to the supplemental file for a detailed description.

The advantages of this strategy are three-fold: 1) it is low-cost and convenient for rapid development iteration, 2) it produces systematic and diverse data, and 3) the data quality is manageable and assessable.
This strategy allows to use the researcher's intuition of the problem space to push the recognition envelope to the fullest degree, for example, by including the training data with different grip gestures and forces.
The risks of this strategy include the potential lack of diverse tapping patterns that a single person can not fully represent as well as the impact of the biomechanics. We mitigate the risks by  evaluating on a separate multi-person dataset. 


\begin{table}[t]
  \centering
  \begin{tabular}{ p{1.8cm}||p{.5cm}||p{4.7cm}}
    \toprule
    {\small \textit{Condition}} & {\small \textit{\#Opt.}} &{\small \textit{Option}}\\
    \cmidrule(r){1-3}
    {\small Phone size} & {\small 2}& {\small Phone A (5.5"), Phone B (6.3")}\\
    {\small Phone case}& {\small 2} & {\small with case, without case}\\
     \cmidrule(r){1-3}
    {\small Orientation} & {\small 4} & {\small portrait, portrait downward, landscape +90$^{\circ}$,  landscape -90$^{\circ}$}\\
     {\small Grip and tap manner} & {\small 7} & {\small 
     tapping with one hand (the four combinations of holding in left/right hand and tapping in left/right hand), both hands with thumbs, tapping on the phone which is lying on a solid surface, and on a soft one.}\\
     {\small Grip force} & {\small 3} & {\small phone rests on the palm, grasp the phone with normal force, grasp with strong force.}\\
     {\small Grip location} & {\small 5} & {\small bottom, mid-bottom, middle, mid-top, top }\\
     {\small Grip gesture} & {\small 4} & {\small grasping the phone against to the palm with one, two, three, and four fingers}\\
     {\small Thumb gesture} & {\small 2} & {\small resting on the screen edge, hovering above the screen}\\
    \cmidrule(r){1-3}
    {\small Tap event} & {\small  2} & {\small tap, non-tap}\\
    {\small Tap finger part} & {\small  2} & {\small finger pad, finger nail}\\
    {\small Tap direction} & {\small  6} & {\small front, back, left, right, top, bottom}\\
    {\small Tap location} & {\small  35 } & {\small 5x7 grid on the housing of the phone}\\
    {\small Tap force} & {\small  5 } & {\small extremely gentle, gentle, normal, strong, extremely strong}\\
    \bottomrule
  \end{tabular}
    \caption{The one-person dataset contains  samples in different conditions of phone, grip gestures, and tapping actions.}~\label{tab:single-person}
    \Description{A table of the data collection conditions for the one-person dataset.}
\end{table}

We repeated data collection under the same condition to mitigate bias. 
In the end, the one-person data collection took over 30 sessions (around half an hour for each) over 69 days.
In total, the one-person training dataset contains 109,200 tapping actions, including 45,500 front taps, 48,450 back taps, 4,020 left taps, 4,020 right taps, 3,610 top taps, 3,600 bottom taps.
Among these tapping actions, 94,764 were collected from a Phone A  and 14,440 from a Phone B, and 85.7\% of them are tapping with finger pad and the rest are with finger nail. 
We also collected non-tap motions (26,060 in total) in a number of scenarios: grasping (4,700) the phone, rubbing the phone case (6,000), releasing a tap (6,160), knocking on the phone placing surface (4,400), shaking and moving with the phone in pocket and in bag (4,800).
The usage of this dataset (135,260 samples including tap and non-tap) was solely in training (or ``teaching''). None was used in any testing to ensure the validity and generalizability of the efficacy measures reported later. 

\subsection{Multi-Person Testing Dataset}

\begin{figure}[t]
\centering
  \includegraphics[width=1.\columnwidth]{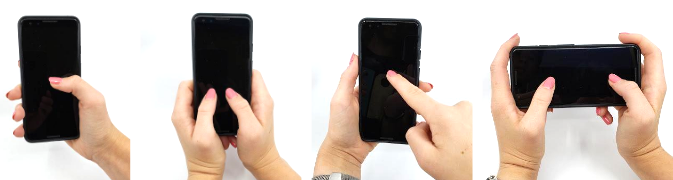}
  \caption{Four common phone grip gestures investigated in the multi-person dataset. Only natural tapping actions with these grip gestures were collected for testing.
  \Description{Four common phone grip gestures.}
  }~\label{fig:grips}
\end{figure}

To evaluate TapNet performance, we collected a multi-person (n=31) dataset.
We focused on collecting the natural tapping actions to simulate tapping data distribution in real use, in particular, with the common finger (thumb and index), in the finger comfort range, and for four most common one-handed and two-handed gestures~\cite{hoober2013users,le2018fingers} as shown in Figure~\ref{fig:grips}.

\subsubsection{Data collection design}


The four grip gestures we studied are in a combination conditions of phone, grip gestures, and tapping actions as shown in Table~\ref{tab:multi-person}.
To keep data collection natural, we avoid continuous and extensive tapping. 
For every condition, participants were asked to tap for five to ten times and take rest at various intervals to avoid fatigue.
In addition, data collection was informed by research conducted on the most common hand grips and finger placements~\cite{hoober2013users,le2018fingers}. 
Unrealistic reach for targets on the back surface of the phone were excluded. 


\begin{table}[t]
  \centering
  \begin{tabular}{ p{2.cm}||p{5.5cm}}
    \toprule
    {\small \textit{Condition}} & {\small \textit{Option}}\\
    \cmidrule(r){1-2}
    {\small Phone size} & {\small Phone A (61\%), Phone B (39\%)}\\
    {\small Phone condition} & {\small with case (58\%), without case (42\%)}\\
     \cmidrule(r){1-2}
    {\small Grip and tap manner} & {\small one-handed/portrait (thumb \& index), two-handed/portrait (thumbs \& indexes), two-handed/landscape (thumbs \& indexes), two-handed/portrait (index)}\\
    {\small Handedness} & {\small right (90\%), left (10\%)}\\
    {\small Hand size} & {\small small (32\%), medium (48\%), large (20\%)}\\
    \cmidrule(r){1-2}
    {\small Tap direction}  & {\small front (54\%), back (28\%), left (4\%), right (4\%), top (5\%), bottom (5\%)}\\
    {\small Tap location}  & {\small comfort range in the 5x7 grid of the screen/back}\\
    \bottomrule
  \end{tabular}
    \caption{The multi-person dataset contains  samples in different conditions of phone, grip gestures, and tapping actions. The number in the parenthesis shows the data percentage.}~\label{tab:multi-person}
    \Description{A table of data collection conditions for the multi-person dataset.}
\end{table}

\subsubsection{Apparatus}

We used a 5.5" Phone A and a 6.3" Phone B throughout the collection. 
To provide the most natural feeling, we assigned phone size in the collection based on participants'  personal phone size.
To achieve a balance ratio, we selected participants according to their personal phone size in recruitment.

\subsubsection{Participants}
We recruited 31 participants (13 female, age: 18-54).
We marked down a number of factors that may affect tapping signal responses, including finger length, finger nail length, hand size, and handedness.
During data collection, the experiment interface would adapt to the handedness of individual participant to ensure the comfort range was valid.

Overall, the multi-person dataset contains 38,545 taps, including  20,615 taps from front, 10,505 back, 1,705 left, 1,705 right, 1,975 top, and 2,040 bottom, among which 58.3\% was collected on phones with rubber cases, and the rest without.

\section{Machine Learning Experiments}

This section evaluates the overall performance of TapNet, followed by the comparison between one- and multi- channel CNN as well as ablation studies on the MIMO network design.

We measured classification performance by F1 score and regression by mean absolute error (MAE) and r$^2$ score. 
F1 score is an weighted average metric of precision and recall, ranging from [0, 1].
The MAE of tap location, i.e. $\sqrt{[(g_x-t_x)/w]^2+[(g_y-t_y)/h]^2}$, is computed by the Euclidean distance between  ground truth location $(g_x, g_y)$ and TapNet output $(t_x, t_y)$ normalized by screen width $w$ and height $h$.
Its range is [0, 1.414], and a baseline (always predicting the center) is 0.707.
Similar to standard deviation, r$^2$  measures how well the regression line fits the data with a range of [0, 1].

In the implementation, training to recognize the presence of tap event requires tap and non-tap data (e.g. phone shaking and rubbing motions).
As non-tap data does not have annotations about tap finger part, direction, and location, it cannot be used to train the rest of the network.
As such, the two parts of the network (tap event v.s. the rest of the tap properties) were trained in turn. 
The tap event branch was trained once after every ten epochs of training the rest of the tap property branches.

We applied ReLU and batch normalization after each convolutional layer and used Adam optimizer with a learning rate of 1e-4 and a momentum decay (1e-6). 
Each model was trained with sufficient number of epochs until the validation (5\% data) loss converges, and training was done using Tensorflow~\cite{abadi2016tensorflow} on a 4G memory GPU. 

The datasets were collected on Phone A and Phone B that ran on Android 9, and the IMU sampling rate was approximately 416Hz.
In run time, it took 0.56 ms on the Phone B main CPU to finish one TapNet inference.
A simplified TapNet can also be run in real time (9ms) on the embedding DSP using Tensor Lite for microcontroller~\cite{tflm2020}, but this is beyond the scope of this paper.
The major latency (105 ms) of the whole pipeline lies in waiting to observe the complete tap signal in the feature window. 

Due to space limitation, we have put some of the implementation evaluations into the supplemental file.
Interesting findings include: 1) high sensor sampling rate is crucial to tap recognition accuracy; 2) the use of gyroscope in our task configuration is important; 
and 3) data augmentation by temporally shifting the samples conduces to the performance gain but scaling the samples does not. 

\subsection{Performance on Tap Recognition Tasks}

This section evaluates TapNet's improvement over prior art and user generalizability when training on one-person data.

\subsubsection{Improving the state of the art}

We implemented and trained four ML algorithms,  including TapNet, to compare their relative performance. The overall performance measured by the weighted average F1 score and MAE across participants and devices are shown in Table~\ref{tab:vs_prior}.
Support Vector Machine (SVM) represents a line of studies~\cite{ling2018sectap,zhang2016beyond} that used traditional machine learning methods. 
TinyCNN is a replication of Liang et al.'s two-layer CNN~\cite{liang2018deep}.
SISO refers to a truncated single-input and single-output (SISO) TapNet. 
It uses the same training configuration (Adam optimizer and learning rate) as the MIMO TapNet, and it gives performance reference if TapNet is configured for a single task.
All the methods were trained on the one-person dataset and tested on the multi-person dataset.

\begin{table}[ht]
  \centering
  \begin{tabular}{ p{1.5cm}||p{.7cm}|p{.8cm}|p{.9cm}|p{1cm}||r}
    \toprule
    \multicolumn{1}{c}{} & \multicolumn{4}{c}{\small{F1 score}} 
    & {\small{MAE (r$^2$)}} \\
    \cmidrule(r){2-6}
    {\small\textit{Method}}
    & {\small \textit{2-class Event}}
    & {\small \textit{2-class Finger part}}
    & {\small \textit{6-class \newline Direction}}
    & {\small \textit{35-class Location}}
    & {\small \textit{Location regression}}\\
    \cmidrule(r){1-6}
    \small{SVM~\cite{ling2018sectap,zhang2016beyond}} & .73 & .93 & .55 & .13 & .20 (.28) \\
    \small{TinyCNN~\cite{liang2018deep}} & \textbf{.89} & .93 & .47 & .03 & .40 (-2.4) \\
    \cmidrule(r){1-6}
    \small{SISO TapNet} & .88 & .89 & .82 & .30 & .19 (.12) \\
    \small{\textbf{TapNet}} & .87 &  \textbf{.94} & \textbf{.85}  & \textbf{.34} & \textbf{.14 (.49)} \\
    \bottomrule
  \end{tabular}
  \caption{Weighted average F1 score of four classification tasks (no color shading) as well as MAE and r$^2$ score of the tap location regression task (purple shading). SVM and TinyCNN are our re-implementation of related works~\cite{liang2018deep,ling2018sectap,zhang2016beyond}. SISO TapNet is the truncated single-input and single-output variant. TapNet is the proposed MIMO method. Note that we have built one model per task for the single-output models. In contrast, TapNet is a multi-task network that gives predictions for all five tasks at a time. Finger part, direction, and location are not conditional on the event detection. The numbers are the higher the better for F1 and r$^2$ scores, while lower the better for MAE.}~\label{tab:vs_prior}
  \Description{Performance comparison table against related work.}
\end{table}

Overall, MIMO TapNet significantly outperforms the prior art~\cite{liang2018deep,ling2018sectap,zhang2016beyond}. 
Compared with the best performance among SVM and TinyCNN, TapNet achieved considerable improvements on tap direction (by 51.0\%) and location (classification: 161.5\% and regression: 30\%).
The SISO TapNet variant also outperforms related works by a marked margin.
TapNet generally outperforms its SISO variant, implying that the MIMO architecture also contributes to performance improvement in addition to the computation and memory benefits.

\begin{figure}[b]
\centering
  \includegraphics[width=1.\columnwidth]{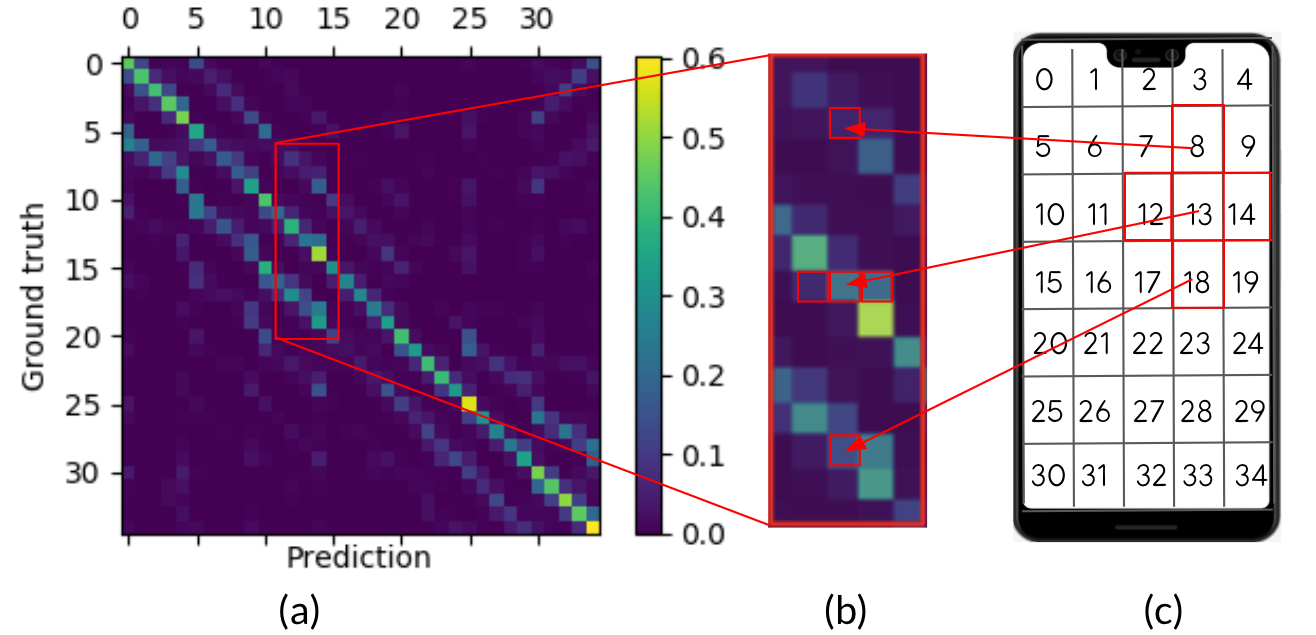}
  \caption{(a) Normalized confusion matrix of the 35-class location classification while training on  one person  and testing on  multiple. (b) An enlarged part of the confusion matrix. (c) The region IDs in a five-by-seven grid. Each region refers to a cell in the grid. Neighboring areas above or below the target has an index offset of five. Predicting to the nearby areas therefore leads to three parallel lines along the diagonal in the confusion matrix. }~\label{fig:confusion}
  \Description{Confusion matrix of the 35-class tap location classification.}
\end{figure}

A close examination of the 35-class tap location classification reveals that TapNet can be usable despite its seemingly limited F1 score (0.34).
Figure~\ref{fig:confusion}(a) shows the normalized confusion matrix of tap location classification. 
Neighboring region above or below the target in a 5x7 grid has an index offset of five.  
The three parallel lines along the diagonal with a five-cell offset in the confusion matrix indicates TapNet either predicts correctly or to nearby regions (see Figure~\ref{fig:confusion}(c)). 
This is in good agreement with the regression results (MAE:.14; $\sim$10\% of the screen diagonal). 
Such location error is similar to the distance between two icons on the phone home screen (see the supplemental video figure).
It thus indicates that IMU-based tap location recognition can be viable and useful in  situations which does not require very high resolution and when capacitive sensing is inadequate (e.g. wearing gloves or under water).

However, the performance difference on simple tasks (event and finger part classification) among different methods is marginal. 
This suggests that the minimal model capacity can be task-dependent; simple tasks may not be benefited from a model capacity increase, but the more complicated tasks can be.

\subsubsection{Achieving person generalizability with one-person data}

We hypothesized that training on diverse but one-person data can still achieve generalizability for unseen users in some tap recognition tasks.
To verify this, we performed evaluation in two paradigms:

\begin{itemize}
	\item \textit{One-to-n-participant} evaluation: TapNet was trained on the one-person dataset and tested on the multi-person dataset. This is the same as the previous evaluation.
	\item \textit{Leave-one-participant-out} cross-validation: TapNet was pre-trained on the one-person dataset, fine tuned on n-1 participants from the multi-person dataset, and tested on the leftout participant in an iterative fashion.
\end{itemize}

By comparing the one-to-n-participant with the leave-one-out evaluation (see  Table~\ref{tab:generalize}), we see that training on a one-person data can still be effective.
As expected, fine tuning on the n-1 participants further improved the performance, but some of the improvements are relatively moderate, for example, by 1\% for tap event and 3.2\% for finger part classification.
This corroborates the hypothesis that training on the diverse data even from a single person can achieve viable user generalizability for tasks, such as event (F1:.92), finger part (F1:.93), and direction classification (F1:.85).
That said, we also learned that tap location recognition relies on very subtle IMU responses, which may relate to personal biomechanics and are hard to simulate by a single person.

\begin{table}[h]
  \centering
  \begin{tabular}{ p{1.6cm}||p{.5cm}|p{.8cm}|p{1cm}|p{1cm}||r}
    \toprule
    \multicolumn{1}{c}{} & \multicolumn{4}{c}{\small{F1 score}} 
    & {\small{MAE (r$^2$)}} \\
    \cmidrule(r){2-6}
    {\small\textit{Evaluation Paradigm}}
    & {\small \textit{2-class Event}}
    & {\small \textit{2-class Finger part}}
    & {\small \textit{6-class \newline Direction}}
    & {\small \textit{35-class Location}}
    & {\small \textit{Location regression}}\\
    \cmidrule(r){1-6}
    \small{one-to-n} & .92 & .93 & .85 &.42 & .15 (.52) \\
    \small{leave-one-out}& \textbf{.93} & \textbf{.96} & \textbf{.92} & \textbf{.54} & \textbf{.11} (\textbf{.73})\\
    \bottomrule
  \end{tabular}
  \caption{Performance comparison between the one-to-n-participant and leave-one-participant-out evaluation. 
  The columns with no color shading are classification tasks, and the one with purple shading is a regression task.}~\label{tab:generalize}
  \Description{Performance comparison table of two evaluation paradigms.}
\end{table}

\subsection{Ablation Studies: Multi- Input and Output}

This section presents the ablation studies on the TapNet architecture: the use of multi-task learning (i.e. multi-output) and cross-device training with auxiliary information (i.e. multi-input).
We evaluated on tap direction classification as a representative task.

\subsubsection{Multi-task learning helps training with limited data}

\begin{figure}
\centering
  \includegraphics[width=1.\columnwidth]{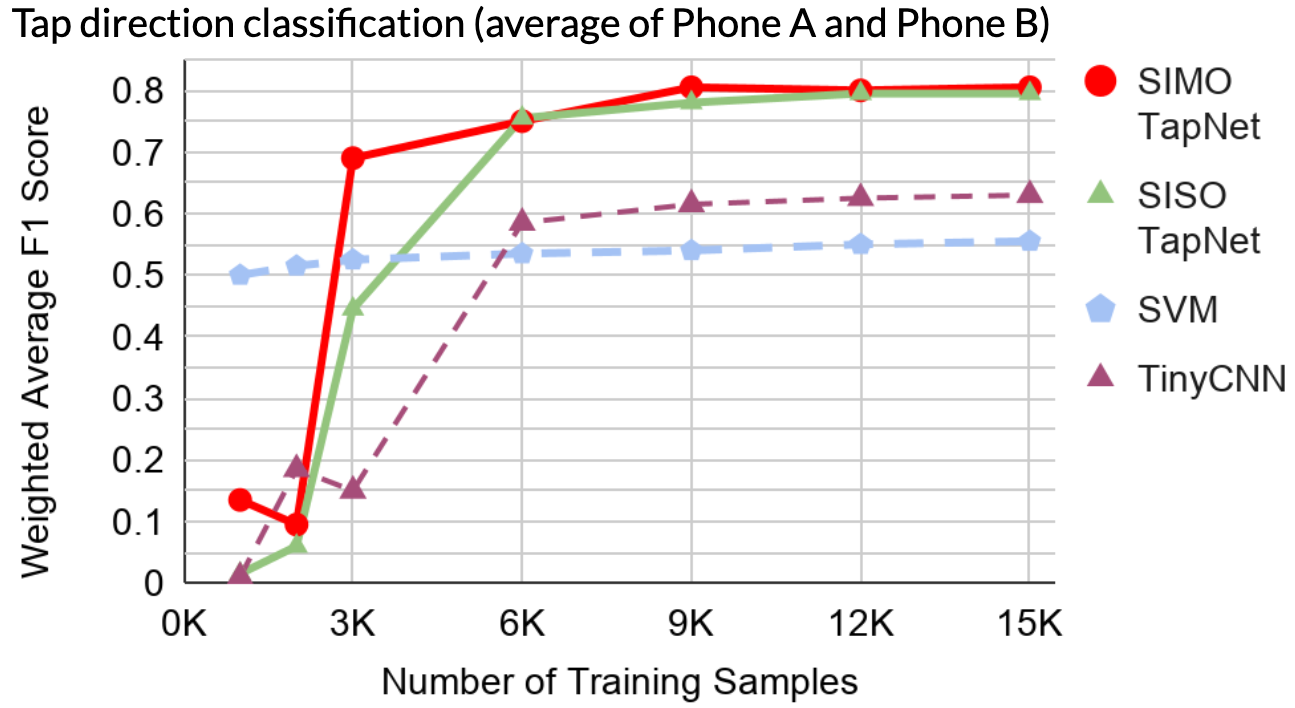}
  \caption{Weighted average F1 score of tap direction classification using multi-task learning (SIMO TapNet) and single-task learning (SISO TapNet). TapNet (either SIMO or its truncated variant) can considerably outperform a SVM~\cite{ling2018sectap,zhang2016beyond} and a tiny CNN model~\cite{liang2018deep}. TapNet with multi-task learning shows advantage given small training data (e.g. 3K samples).
  }~\label{fig:multi-output}
  \Description{A figure of performance comparison for multi-output evaluation.}
\end{figure}

We first present the evaluation on multi-task learning. 
We compared against the single-output counterpart (SISO; a truncated TapNet variant) and recent related works, including a SVM~\cite{ling2018sectap,zhang2016beyond} and a tiny CNN model~\cite{liang2018deep}.
Limited by the tap samples on Phone B ($\sim$15K) in the training dataset, we evaluated model performance averaged over two devices with incremental training samples from 1K to 15K. 
Figure~\ref{fig:multi-output} shows the comparison results.

Multi-task learning contributes to tap direction classification, especially with a small amount (no more than 3K) of training samples.
Nevertheless, with sufficient training samples (over 3K) performance of SIMO and SISO TapNets flats out with a comparable F1 score (SIMO:.81, SISO:.80). 
Although the performance improvement of SIMO over SISO is modest in this specific task, its advantages of reusing computation (for the shared layers) across recognition tasks and memory saving to avoid running multiple models on small processing units are essential. 
One the other hand, both SIMO and SISO TapNets can considerably outperform TinyCNN (purple dashed) starting from 3K training samples, and outperform SVM (blue dashed) starting from 6K.
Taken together, a multi-output TapNet is favorable over the state of art~\cite{liang2018deep,ling2018sectap,zhang2016beyond}.

\subsubsection{Cross-device training utilizes data efficiently}

\begin{figure}[t]
\centering
  \includegraphics[width=1.\columnwidth]{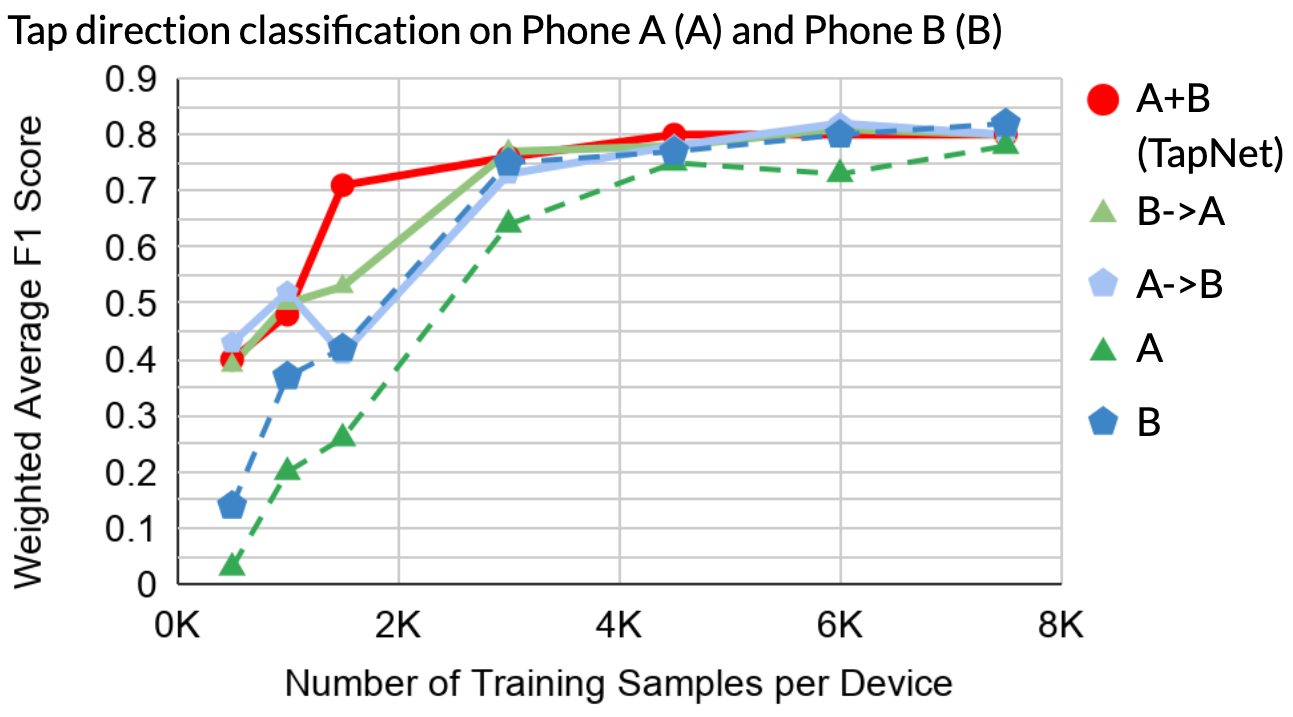}
  \caption{Performance comparison of tap direction classification with auxiliary information (A+B) against fine tuning (B->A, A->B), and training on device-specific data (A, B) alone. X-axis shows the number of training samples per device and y-axis indicates the weighted average F1 score. The jointly trained TapNet yields comparable performance to that of the fine tuned models on specific devices and outperforms the models trained on device-specific data alone. }~\label{fig:multi-input}
  \Description{A figure of performance comparison for multi-input evaluation.}
\end{figure}

To investigate cross-device model training, we evaluated model performance with incremental training samples from 1K to 15K, as previous experiments suggest that TapNet converges with around 15K training samples.
Note that in the case of training with 15K samples, TapNet was jointly trained on 7.5K samples from Phone A and another 7.5K samples from Phone B.
Similarly, its counterparts were pre-trained on 7.5K samples from one device and fine tuned on the other.
As baselines, we also evaluated training on the 7.5K device-specific samples alone for each device.

Figure~\ref{fig:multi-input} gives the performance comparison. 
A+B represents TapNet jointly trained on cross-device data; B->A and A->B denote tuning the pre-trained model from one device on the other; and A and B indicate models trained on device-specific data.
The overall performance of different models increases with training samples and they flats out after 4.5K samples per device. 
More interestingly, cross-device training with sensor location as auxiliary information (A+B) can approach the performance upper bound earlier (with 1.5K samples per device) than the rest of the models.

\subsection{Signal Alignment in One-Channel CNN}

To verify the efficacy of one-channel CNN for tap recognition, we performed a comparison on tap direction classification, which is a representative task and demonstrated to require subtle signal alignment across channels.
As input data dimension affects the number of trainable parameters (i.e. model capacity) of a specific network architecture, we evaluated similar architectures with commensurate number of trainable parameters, so as to compare models with comparable capacity.
This experiment was conducted on the Phone A data using the one-to-n evaluation paradigm.

\begin{figure}[t]
\centering
  \includegraphics[width=1.\columnwidth]{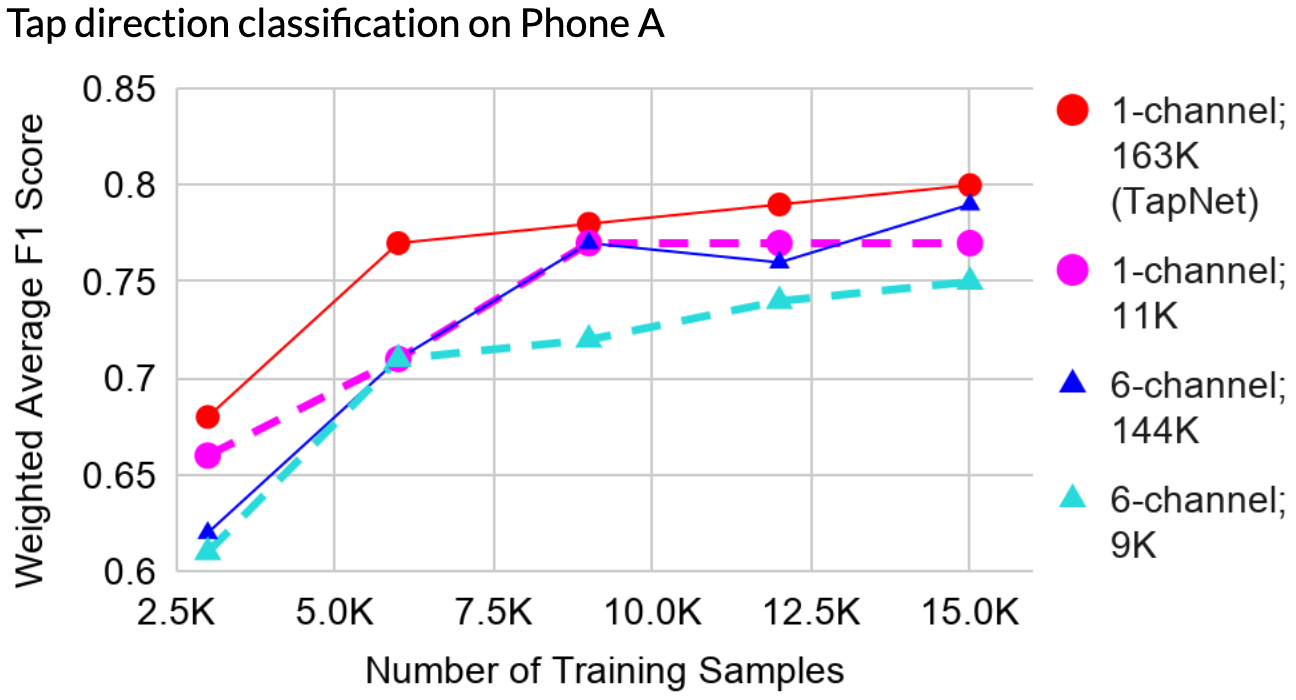}
  \caption{Performance comparison of one-channel CNN (TapNet) against its six-channel counterpart with two levels of model capacity (large models: 163K, 144K trainable parameters; small models: 11K and 9K). These are the average results over three runs to cancel out training randomness. The one-channel CNN with 11K parameters can achieves comparable performance as that of the six-channel CNN with a larger number of parameter (144K). We tuned the number of filters and maintained its ratios across layers to roughly match the total number of trainable parameters between the one-channel and six-channel models.}~\label{fig:1c6c}
  \Description{A figure of performance comparison for one-channel CNN evaluation.}
\end{figure}

Figure~\ref{fig:1c6c} shows the performance comparison of one-channel CNN against its multi-channel counterparts.
The x-axis shows the number of training samples, and the y-axis the weighted average of F1 score.
The solid lines denote the performance of large-capacity models with 163K (one-channel) and 144K (six-channel) trainable parameters, while the dashed lines those of small-capacity models with 11K (one-channel) and 9K (six-channel) trainable parameters.
Most importantly, small-capacity one-channel CNN (purple dashed) and large-capacity six channel CNN (blue) have almost equivalent performance with 6K-12K training samples. 
Further, comparing the solid and dashed lines, large-capacity models generally outperforms small-capacity models with similar architecture and input format.
Taken together, we conclude that the one-channel CNN can address across-channel signal alignment more efficiently than its multi-channel counterpart for tap property recognition.

\section{Discussion}

We designed, developed, and evaluated a set of deep learning methods for on-device IMU signal based off-screen input, in particular TapNet, a multi-task network that allows for cross-device training with phone form factor as auxiliary information and joint prediction of tap-related tasks, including tap event, direction, finger part, location classification, and the regression of tap location. 
This architecture not only shares knowledge across tap properties and devices during training, but also shares computation and memory at run time.
Some of the TapNet building blocks are not novel in fields such as computer vision, but to bring them into building an interaction gesture to a practical performance level required original research.
TapNet achieved a marked improvement over the prior art~\cite{liang2018deep,ling2018sectap,zhang2016beyond} on tap direction (by 51.0\%) and location (161.5\%) classifications as well as tap location regression (30\%).

We also discovered encouraging  generalizability across users when the model was trained (or "taught") by a one person.
We have tested TapNet on those who had not been in any set of the data collection and encouraged them to explore the effects of TapNet freely on their own device in daily use (for such functions as taking screenshot). Their experience matched with the leave-one-participant-out test results reported here. 
This high generalizability is probably because the inter-class differences (e.g. finger pad vs. nail in the finger part classification task) are generally much greater than the inter-person differences, such that the personal biomechanics only imposes a neglectable effect. 

\begin{figure}[t]
\centering
  \includegraphics[width=1.\columnwidth]{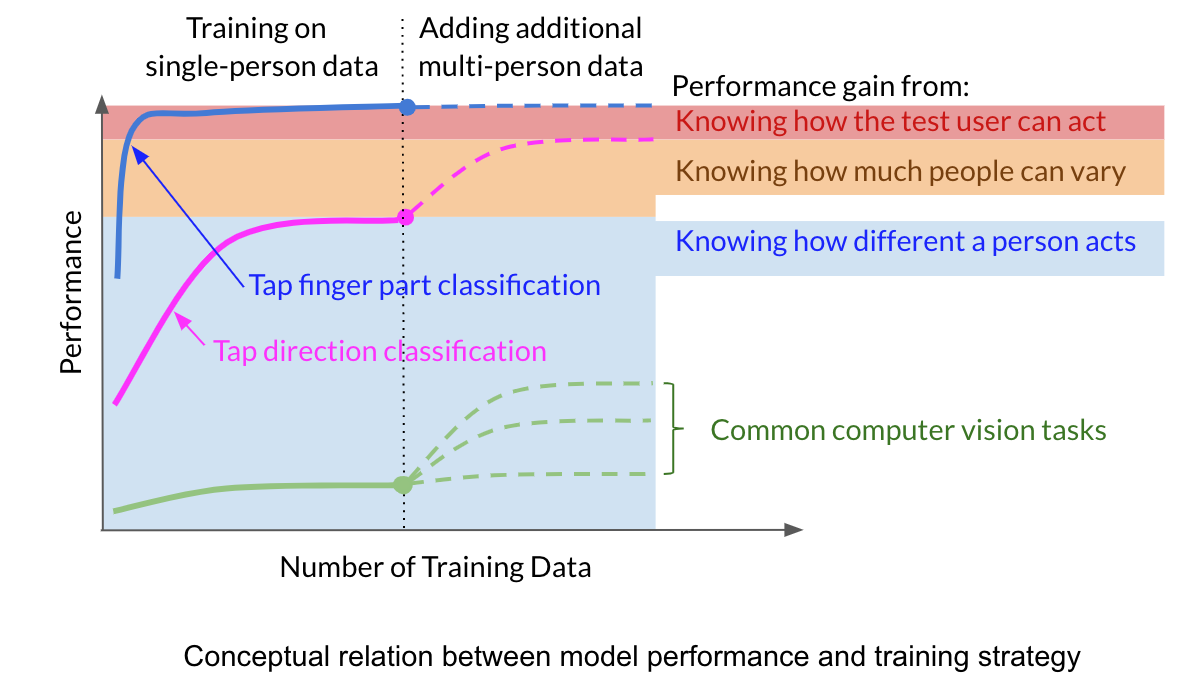}
  \caption{Conceptual relation between  performance and training strategy with one-person and multi-person data.}~\label{fig:conceptual}
  \Description{A figure of conceptual relation between model performance and training strategy.}
\end{figure}

The other advantage of our approach is that an expert design could intentionally push the variations of an intentional tap gesture in terms of speed, strength, angle, and hand posture, 
However, it is possible certain recognition tasks, such as higher resolution tap location classification, could demand more fine-grained information only available in person-specific data.  Training on incremental one-person data can increase performance up to a certain level and then plateaus (see Figure~\ref{fig:conceptual}).
Further adapting the model to multi-person data may teach the model to understand the artifacts of personal biomechanics, and thus further improves the performance until its next plateau.
This performance gain is from knowing how much people can vary.
To reach the ideal performance, it becomes a must to know the person-specific information, i.e. how exactly the target user acts.
In practice, especially for learning-based gesture studies, it can be beneficial to identify the curve in Figure~\ref{fig:conceptual}, and then decide the needed number of training participants for specific tasks.

TapNet opens up new interaction opportunities such as one-handed interaction  that uses off-screen tapping or designated tap gestures. 
TapNet is robust to phone case and fabric, and thus can be used for wearable interaction without specialized smart fabrics~\cite{dobbelstein2017pocketthumb}. 
This study also offers new potential for other research domains, including biometrics.
For instance, it is possible to improve continuous and passive authentication~\cite{amini2018deepfp,xu2014towards,zheng2014you} by using tap properties that contain biometric information, i.e. with a clear gap between the first two plateaus (orange area) in Figure~\ref{fig:conceptual}.



Although evaluation results demonstrated that TapNet benefited from cross-device training, we do not expect the current TapNet trained on these two devices would directly apply to unseen devices without further training, i.e. a cross-device model. However, the joint training architecture has been shown effective and this is a promising step toward a device adaptive model.
We have set up the infrastructure and we plan to further investigate along this line by adding new devices and also opensource our implementation.

\section{HCI applications of TapNet}

Without adding new sensor hardware, TapNet enables many new input possibilities on smartphones. In addition to providing shortcuts to quick activations of apps or functions such as camera and screenshot, this section sketches out four applications enabled by TapNet's capability of measuring multiple proprieties in addition to the presence of tap event, such as direction and location.
Please refer to our supplemental video for these applications in action.

\begin{figure}[t]
\centering
  \includegraphics[width=1.\columnwidth]{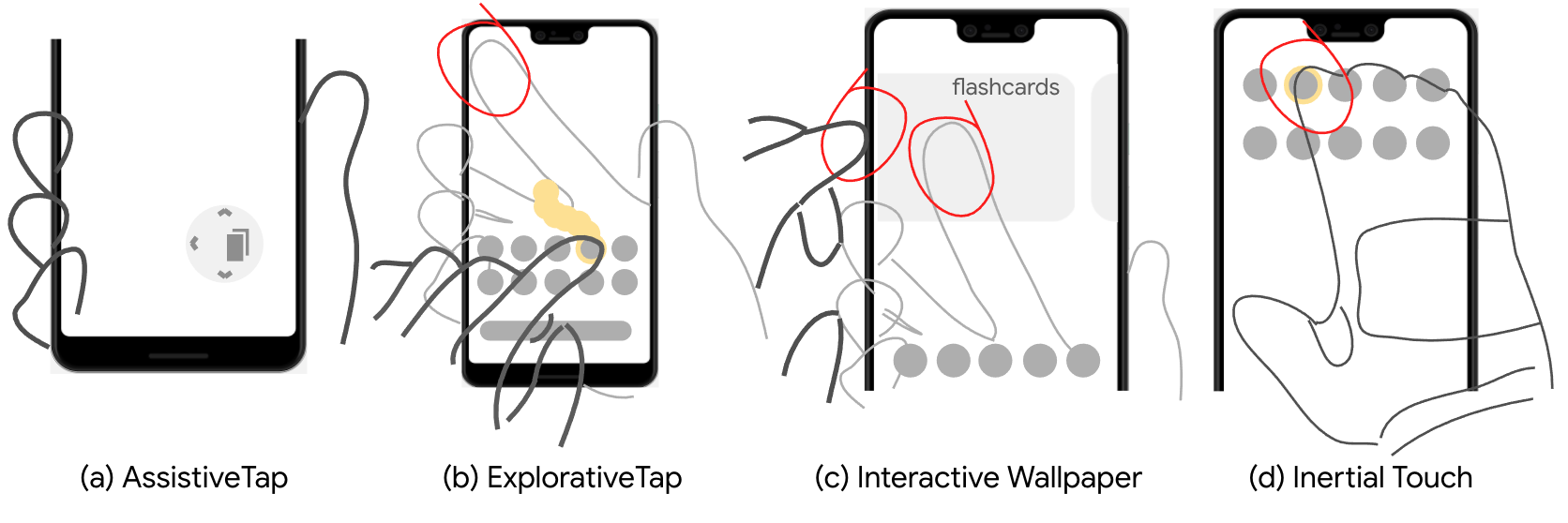}
  \caption{Four example use cases enabled by TapNet. AssistiveTap uses back tap and tilting for one-handed interaction. ExplorativeTap introduces a new two-sided interaction paradigm for users with visual impairments. Interactive wallpapers enables interaction with UI background objects. Inertial Touch senses tapping by force and angle changes and provides auxiliary information for capacitive sensing.}~\label{fig:applications}
  \Description{A figure of four example use cases of TapNet.}
\end{figure}

\subsection{AssistiveTap}

AssistiveTap allows users to complete a number of system interactions using back tap and tilting.
Users can perform a back double tap to invoke the AssistiveTap interface (see Figure~\ref{fig:applications}a), then tilt the phone (based on the same IMU signal input to TapNet) to select gesture, and back tap to perform the selected gesture (e.g. 'back' gesture, scrolling, app switch).
In other words, AssistiveTap provides the back-of-device alternatives for the most commonly used on-screen gestures, by using the gesture combination of back tap and tilting.
It can be beneficial in situations, where one-handed interaction is preferable or even required.


\subsection{ExplorativeTap}

ExplorativeTap is a two-handed interaction method that exploits signals from both the front (touch) and back of the phone tap (see Figure~\ref{fig:applications}b). 
Users can use the exploration finger on the screen to glide over the on-screen objects, hear them, and perform a back tap to confirm selection.
This is an improvement for the visual impairment accessibility modes, such as Voice Over on iOS and  Talk Back on Android.
These accessibility modes occupy the common on-screen gestures (e.g. swipe and touch), and their users need to learn a more complicated system navigation gestures. 
In contrast, ExplorativeTap starts the accessibility mode by a double back tap, selects object by a single back tap, and exits by lifting the exploration finger.
It, therefore, solves the conflicts with system navigation gestures, and thus can be helpful for users with low vision and print disability, who need quick and temporary access to the accessibility mode.




\subsection{Interactive Wallpaper}

Off-screen tap recognition makes it possible to interact with objects living in different interface layers, such as background targets that do not react to on-screen touch.
For instance, TapNet can enable  users to interact with flashcards or news feeds shown in the wallpaper (see Figure~\ref{fig:applications}c). 
They can back tap to change the flashcard or news feeds, or edge tap to switch a different set of them, and these can be done even on the lock screen.
It  enables the user to utilize the bits and pieces of time for their favorite spare time activity. 

\subsection{Inertial Touch}

Inertial Touch estimates tap force from the IMU signals and tap location from the TapNet output. 
We can define Inertial Touch as \textit{a fast touch event with a strong tapping momentum}.
As TapNet estimates tap location from force and angle changes, it can function even when capacitive sensing fails (Figure~\ref{fig:applications}d). 
Common use situations include when users are wearing gloves, having long fingernails, or when there are waterdrops on the touchscreen.


\subsection{Summary of Use Cases}

The aforementioned use cases can be particularly useful in challenging situations when one-handed and non-contact interactions are preferable or required.
For example, AssistiveTap exploits back tap and titling to partially address the issue of limited thumb reaching area during one-handed interaction. 
ExplorativeTap allows for the coordination between on-screen and off-screen interactions and thus saves the need for learning additional set of on-screen gestures. 
Inertial Touch offers auxiliary impact information beyond capacitive sensing.
These are just a few application examples of tapping into TapNet's ability to detect multiple tap properties.

\section{Conclusion}

This paper presents the design, training, implementation and application of TapNet for off-screen mobile input. 
TapNet employs a multi-task convoluational neural network with motion signals as primary input and phone form factor as auxiliary information.
It allows for joint learning on data across devices and simultaneous estimation of multiple tap properties. 
In comparison to many alternatives,  this neural network  architecture worked the best towards our goal of increasing the UI design space of off-screen interactions.
To train and optimize this and other alternative ML models, we developed a one-person dataset for training and a multi-person dataset for testing.
The evaluation results show that 1) TapNet significantly outperformed the state of the art especially in difficult recognition tasks such as tap location estimation; 2) multi-task learning is more data efficient, showing greater advantage particularly with a limited amount of training data; 3) cross-device training with phone form factor increases the efficiency of data utilization; and 4) one-channel CNN can achieve cross-channel signal alignment more efficiently than its multi-channel counterpart.
We verified the hypothesis that training on the one-person data can still generalize well across users if the diversity of the training data can be ensured.
This sheds light on the conceptual relation between model performance and ML training strategy in IMU-based input systems.
We demonstrated that many new interaction use cases could be enabled by TapNet.
Taken together, the TapNet project made significant progress towards practically enabling and enlarging the off-screen interaction design space by deep learning from on-device IMU signals,  establishing new benchmarks with reproducible results, datasets and codebase.




\bibliographystyle{ACM-Reference-Format}
\bibliography{main}

\end{document}